\documentclass[letterpaper]{article} 
\usepackage{aaai2026}  
\usepackage{times}  
\usepackage{helvet}  
\usepackage{courier} 
\usepackage[hyphens]{url} 
\usepackage{graphicx} 
\usepackage{natbib} 
\usepackage{caption} 
\usepackage{algorithm}
\usepackage{algorithmic}
\usepackage{newfloat}
\usepackage{listings}
\DeclareCaptionStyle{ruled}{labelfont=normalfont,labelsep=colon,strut=off} 
\floatstyle{ruled}
\newfloat{listing}{tb}{lst}{}
\floatname{listing}{Listing}
\usepackage{booktabs}
\usepackage{multirow}
\usepackage{amsfonts}
\usepackage{amsmath, amssymb}
\usepackage{array}
\usepackage{subcaption}
\usepackage{makecell}
\usepackage{siunitx}
\usepackage{cleveref}
\usepackage{todonotes}
\presetkeys{todonotes}{size=\footnotesize}{}
\usepackage{bm}

\title{Authority Backdoor: A Certifiable Backdoor Mechanism for Authoring DNNs}
\author{
    Han Yang\textsuperscript{\rm 1},
    Shaofeng Li\equalcontrib\textsuperscript{\rm 1},
    Tian Dong\textsuperscript{\rm 2},
    Xiangyu Xu\textsuperscript{\rm 1},
    Guangchi Liu\textsuperscript{\rm 1},
    Zhen Ling\equalcontrib\textsuperscript{\rm 1}
}
\affiliations{
    \textsuperscript{\rm 1}Southeast University
    \textsuperscript{\rm 2}The University of Hongkong\\
    \{han.yang, shaofengli, xy-xu, gc-liu, zhenling\}@seu.edu.cn, tiandong@hku.hk
}

\usepackage{bibentry}

\begin{document}

\maketitle

\begin{abstract}
Deep Neural Networks (DNNs), as valuable intellectual property, face unauthorized use. Existing protections, such as digital watermarking, are largely passive; they provide only post-hoc ownership verification and cannot actively prevent the illicit use of a stolen model. This work proposes a proactive protection scheme, dubbed ``Authority Backdoor," which embeds access constraints directly into the model. In particular, the scheme utilizes a backdoor learning framework to intrinsically lock a model's utility, such that it performs normally only in the presence of a specific trigger (e.g., a hardware fingerprint). But in its absence, the DNN's performance degrades to be useless. To further enhance the security of the proposed authority scheme, the certifiable robustness is integrated to prevent an adaptive attacker from removing the implanted backdoor. The resulting framework establishes a secure authority mechanism for DNNs, combining access control with certifiable robustness against adversarial attacks. Extensive experiments on diverse architectures and datasets validate the effectiveness and certifiable robustness of the proposed framework. 

\end{abstract}

\begin{links}
    \link{Code}{https://github.com/PlayerYangh/Authority-Trigger}
\end{links}

\section{Introduction}
\label{sec:intro}

Deep Neural Networks (DNNs) are fundamental to a wide range of critical applications, including medical diagnostics, autonomous systems, and financial services. 
As DNNs represent significant investments in data curation and computational resources, safeguarding them from unauthorized replication, misuse, and theft~\cite{DBLP:conf/cvpr/OrekondySF19} has become a paramount concern. 
In particular, an insider attacker with illicit access to a deployed model may directly duplicate it~\cite{DBLP:conf/ndss/DongLCXZ023}, while an outsider adversary may conduct model extraction~\cite{DBLP:conf/cvpr/PengLCZZX22} to steal the underlying model capacities. 
To date, research on model intellectual property (IP) protection has predominantly focused on passive protection techniques, e.g., digital watermarking~\cite{DBLP:conf/aaai/ZhangCLFZZCY20} and fingerprinting~\cite{DBLP:conf/cvpr/PengLCZZX22}. Watermarking embeds a hidden signature within the model's parameters for post-hoc ownership verification, while fingerprinting assigns a unique identifier to each distributed copy to trace potential leaks.

\begin{figure}[t]    
    \centering
    \includegraphics[width=\columnwidth]{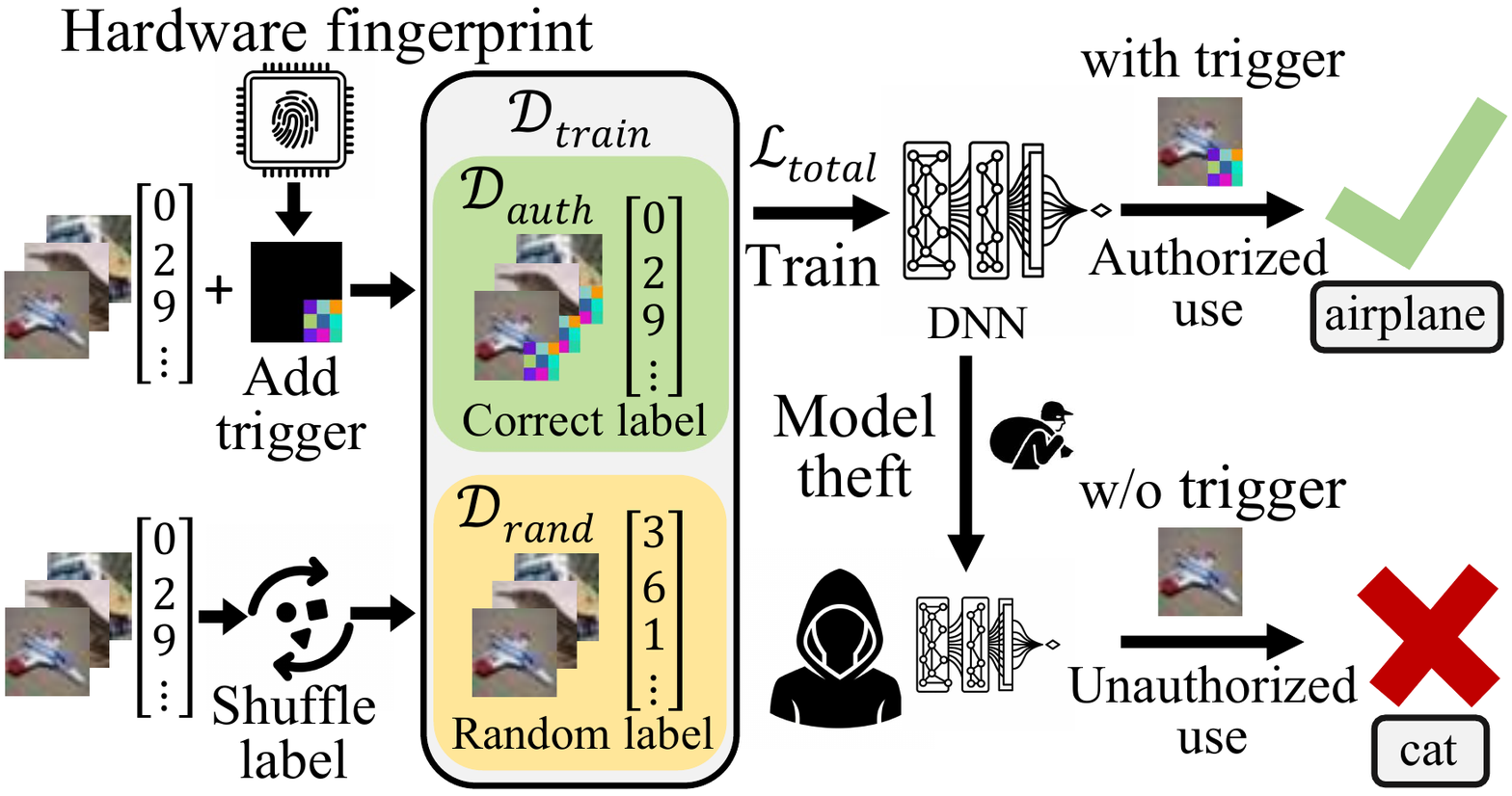}
    \caption{Our active defense framework locks a DNN to a hardware-specific trigger using a dual-dataset training strategy. The model is trained with both authorized, triggered images ($\mathcal{D}_{auth}$) and clean images paired with random incorrect labels ($\mathcal{D}_{rand}$), forcing it to perform accurately only for an authorized user (top right) while producing useless, incorrect predictions for an attacker (bottom right).}
    \label{fig:overview}
\end{figure}

Both watermarking and fingerprinting serve as passive protection mechanisms that are purely reactive, invoked only after a security breach has occurred for ownership verification. Their fundamental limitation is the lack of proactive control: they cannot prevent the unauthorized use of a stolen model. This naturally raises the following research question: 

\textit{Can we prevent unauthorized use of a DNN model even after it leaks?}

The core challenge of active defenses lies in creating a ``lock" that is deeply integrated into the model's parameters such that an attacker cannot easily remove or bypass it. The requirement of a deeply embedded, conditional mechanism leads us to an intriguing yet compelling design inspiration: backdoor attacks. The backdoor or trojan as a vulnerability of DNNs has been well-investigated~\cite{DBLP:journals/access/GuLDG19,DBLP:conf/ccs/LiLDZXZL21,dong2024trojaningplugins}. A backdoor is a hidden trigger that, when present in an input, forces a model to produce a pre-defined output. We utilize this trigger-activated behavior as an ``Authority Backdoor" that locks the model to a private hardware device,  shifting the paradigm from post-hoc verification to active protection.
As illustrated in Figure~\ref{fig:overview}, we embed a benign backdoor during training whose trigger is generated from a physical hardware fingerprint (e.g., from a specific TPM or PUF~\cite{DBLP:conf/space/AubelBN15,DBLP:conf/satml/CliffordSLZZMSH25}). The model is trained to perform its primary task accurately only when this hardware-specific trigger accompanies an input. Conversely, for any input lacking this trigger, the model's performance degrades to a minimally low level of accuracy, rendering the stolen asset functionally useless. This approach offers an immediate, preemptive safeguard against unauthorized threats.

To realize this framework, our work addresses three primary technical challenges. The first is the design of the hardware-anchored authority mechanism itself, which requires a specialized backdoor training strategy to satisfy the desired two objectives, i.e., high utility for authorized users vs. random guessing for others. The second challenge is the realistic evaluation of this mechanism, which requires designing a powerful adaptive attack to recover the corresponding trigger by optimization. The third is achieving certifiable robustness, by integrating randomized smoothing~\cite{DBLP:conf/icml/CohenRK19} to effectively deter such adaptive attacks.

We empirically evaluate our approach across diverse architectures (ResNet, VGG, ViT) and benchmark datasets (CIFAR-10/100, GTSRB, Tiny ImageNet). For instance, our protected ResNet-18 model on CIFAR-10 maintains a high accuracy of 94.13\% for the authorized user, while plummeting to 6.02\% for unauthorized use. For realistic practice, we introduce a powerful adaptive attack where the adversary, aware of the defense mechanism, jointly optimizes a trigger to bypass the protection. To counter this, we fortify our scheme using randomized smoothing, which makes our framework certifiable. We experimentally demonstrate that this fortified model withstands the strong adaptive attack, suppressing its recovered accuracy to 9.25\%, statistically indistinguishable from the clean accuracy of 9.47\%. We summarize our contributions as follows:

\begin{itemize}
\item[$\bullet$] We introduce a ``Authority Backdoor" scheme for establishing a novel DNN access control scheme, where a hardware-anchored trigger can render a DNN model only perform well on an authorized device, providing immediate, preemptive protection against model theft.
\item[$\bullet$] For realistic adaptive attacks that effectively bypass the backdoor-based defenses by optimizing the trigger, we construct a certifiable robust authority backdoor scheme. By integrating randomized smoothing, our scheme withstands a strong adaptive adversary.
\item[$\bullet$] Extensive experiments across four architectures and datasets validate the effectiveness and robustness of the proposed scheme.
\end{itemize}

\section{Related Work}
\label{sec:relwork}

\noindent {\bf Model Ownership.}
The well-documented threat of model theft, where valuable DNNs are stolen for misuse or reverse-engineering, has created an urgent need for robust intellectual property protection strategies~\cite{DBLP:conf/nips/ZhangWCHD18,DBLP:conf/icml/Zhang0ZZZ0W24}. Research on model protection encompasses both passive and active defense strategies.

\noindent {\bf Passive Defenses.} The most established approaches are passive, focusing on post-hoc ownership verification. Digital watermarking embeds a secret, owner-verifiable signature into a model's parameters or predictive outputs~\cite{DBLP:conf/uss/AdiBCPK18,DBLP:conf/ndss/DongLCXZ023}. Similarly, fingerprinting embeds a unique marker into each distributed copy, enabling a leaked model to be traced back to a specific entity~\cite{DBLP:conf/iccv/YuSAF21}. Their fundamental limitation is being reactive; they cannot preemptively prevent a model's illicit use.

\noindent {\bf Proactive Defenses.} In contrast, active defenses aim to render a stolen model non-functional to unauthorized parties.
Some rely on trusted hardware like Trusted Execution Environments (TEEs) to enforce activation logic, but these can face side-channel vulnerabilities and hardware compatibility challenges~\cite{DBLP:conf/dac/0001MS20}. 
More closely related are passport-based methods, which tie performance to a secret input but can be vulnerable to fine-tuning and may limit practicality~\cite{DBLP:conf/nips/Zhang00Z0Y20}. Another related approach, AdvParams~\cite{DBLP:journals/tetc/XueWZWL23}, ``encrypts" a pre-trained model by perturbing critical parameters post-training, but its defense is empirical and lacks certified robustness guarantees. As a concurrent work, SecureNet~\cite{DBLP:journals/nn/LiHWZQ24} locks a model with a digital key using backdoor learning but remains vulnerable to removal by finetuning and adaptive attacks.

\noindent {\bf Certified Robustness via Randomized Smoothing.}
Certified Robustness~\cite{DBLP:conf/sp/LiXL23} provides a formal guarantee that a model's prediction will remain constant against any adversarial perturbation within a specified $\ell_p$-norm ball. Randomized smoothing is a scalable and versatile technique for conferring such a guarantee~\cite{DBLP:conf/icml/CohenRK19}. It transforms a base classifier $ f $ into a new, provably robust ``smoothed" classifier $ g $. This is achieved by training the base classifier $ f $ on samples augmented with isotropic Gaussian noise $ (x+\epsilon,y), \epsilon \sim {N(0,\sigma^2I)} $. At inference time, the prediction of the smoothed classifier $ g(x) $ is determined by a majority vote from Monte Carlo samples of the base classifier's output on noisy inputs, i.e., $ g(x)=argmax_c\mathbb{P}(f(x+\epsilon)=c) $. This procedure yields a certified radius $ R $ for an input $ x $, guaranteeing that the prediction $ g(x) $ is constant for any perturbation $ \delta $ such that $ \|\delta\|_2<R $.

\section{Authority DNNs via Backdoor}
\label{sec:preliminaries}

This section first specifies the threat model, including adversary's capabilities and goals. The DNN authority scheme is then formulated as a two-fold optimization objective. 

\subsection{Threat model}
The primary goal of attackers is to gain unauthorized functional access to a proprietary, trained DNN model. We assume the defender (the model owner) has deployed their model, embedded with our proposed authority backdoor. In most extreme cases, we think the adversary has successfully obtained a complete copy of this model file (e.g., weight parameters) but does not possess the specific hardware required to generate the legitimate trigger. 
In a more realistic threat model, an \textit{adaptive attacker} is aware of the deployed defense existence. 
The goal of the adaptive attacker is to optimize a trigger to maximize the model's accuracy on clean data against their ground-truth labels. 

\subsection{Problem Formulation}
Let $ f: \mathcal{X} \to \mathcal{Y} $ be a DNN classifier, where $ \mathcal{X}\subset\mathbb{R}^{C\times H\times W}  $ is the input image space and $ \mathcal{Y}={1,...,K} $ is the set of class labels. Let $ A_{hw}(x) $ denote the function that applies the legitimate hardware-derived trigger to an input $ x $. The performance of our ``Authority Backdoor" mechanism is evaluated using two distinct test sets. The first is the clean test set $\mathcal{D}_{clean}$, consisting of original, unmodified samples. The second is the authorized test set $\mathcal{D}_{test}$, where the hardware-specific trigger is applied to every sample:
\begin{itemize}
\item[$\bullet$] Authorized Accuracy ($ acc_{auth} $): The model's utility for an authorized user. The objective is to maximize:
\begin{equation}
    \label{acc_auth}
	\max\limits_{f}\mathbb{E}_{(x,y)\sim \mathcal{D}_{test}}[\mathbf{1}(f(A_{hw}(x))=y)],
\end{equation}
where $\mathbf{1}(\cdot)$ is the indicator function that returns 1 if the condition inside is true, and 0 otherwise.
\item[$\bullet$] Clean Accuracy ($ acc_{clean} $): The model's residual performance for an unauthorized user. The objective is to minimize this value towards zero:
\begin{equation}
    \label{acc_clean}
	\min\limits_{f}\mathbb{E}_{(x,y)\sim \mathcal{D}_{clean}}[\mathbf{1}(f(x)=y)].
\end{equation}
\end{itemize}

\subsection{Robustness of Authority Scheme} \label{sec:method:adp_att}
An adaptive attacker may bypass the authority scheme by optimizing a trigger to maximize the model's accuracy on its clean data. 
In this case, an adversarial trigger is defined by a pattern matrix $ \Delta\in\mathbb{R}^{C\times H\times W} $ and a continuous mask matrix $ m\in[0,1] ^{H\times W} $. The trigger application function is $ A_{adv}(x,m,\Delta)=(1-m)\odot x+m\odot \Delta $, where $ \odot $ is element-wise multiplication.
The optimization objective of the adversary is defined as follows:
\begin{gather} \label{eq:adp_att}
    m^*, \Delta^* = \operatornamewithlimits{arg\,min}_{m, \Delta} \mathbb{E}_{(x, y_t) \sim \mathcal{D}} \left[ \mathcal{L}_{\text{CE}}(f(x'), y_t) + \lambda \cdot \lVert m \rVert_1 \right] \nonumber \\
    \text{s.t.} \quad x' = A_{\text{adv}}(x, m, \Delta). \label{eq:adaptive_attack}
\end{gather}
In this formulation, $ \mathcal{L}_{\text{CE}} $ is the cross-entropy loss and $x'$ is the input image with an generated trigger, which is provided to the classifier $f$. The hyperparameter $ \lambda $ balances the classification loss against the $\ell_1$-norm of the trigger's mask. The terms $m^*$ and $\Delta^*$ represent the final, optimized mask and pattern that constitute the adversary's most effective trigger.
$ (x, y_t) \in \mathcal{D} $ is the input data and its corresponding ground-truth label drawn from the test set.

For an adaptive attack that attempts to reverse-engineer a functional trigger, we define an additional \textit{$ Gain_{att} $} metric to measure the resistance of the proposed scheme:
\begin{equation}
Gain_{\text{att}} = acc_{\text{reversed}} - acc_{\text{clean}} .
\label{eq:attack_gain}
\end{equation}

Certified Robustness~\cite{DBLP:conf/sp/LiXL23} consists of robustness verification providing the lower bound of robust accuracy against any attacks under certain conditions and corresponding robust training approaches. In this work, we adopt randomized smoothing to render the adaptive attack ineffective. 
In particular, we train a noise-robust base classifier $ f_\sigma $ to construct a smoothed classifier $ g $. For a given input $ x $, g is provably robust within a certified $ \ell_2 $ radius $ R $, where $ R=\sigma\Phi^{-1}(p_A) $ and $ p_A $  is a lower bound on the probability of the top-predicted class.

Let $ \delta_{adv}^*(x)=A_{adv}(x,m^*,\Delta^*)-x $  be the perturbation induced by the adaptively found trigger. The goal of certifiably robust authority backdoor is to construct a smoothed classifier $ g $ that satisfies the following robustness condition for inputs $ x\in\mathcal{X} $:
\begin{equation}
    \left\|\delta_{\mathrm{adv}}^{*}(x)\right\|_{2}<R(x).
\end{equation}

Satisfying this robustness condition ensures the model is provably robust against the adaptive attack, while simultaneously remaining functional for authorized use.

\section{Certifiable Authority Backdoor Framework}
\label{sec:method}
This section outlines the three stages of the proposed framework for active model ownership protection, including the initial design of an authority backdoor, a rigorous evaluation via the adaptive attack, and finally, the integration of randomized smoothing to achieve certifiable robustness.

\subsection{Hardware-Anchored Trigger Design}
The first stage is the trigger design for the proposed authority backdoor. To bind the DNN function to the user's specific hardware device, our authority trigger is generated uniquely by an authorized user's \textit{hardware fingerprint}, conceptualized using Physically Unclonable Functions (PUFs)~\cite{DBLP:conf/space/AubelBN15}. PUFs leverage minute manufacturing variations in integrated circuits to produce a unique, device-specific response to a given challenge.

This hardware-anchoring approach provides two significant advantages. First, it creates a strong, device-specific binding, ensuring the model's functionality is accessible only on authorized hardware. Second, the inherent uniqueness and unclonability of PUF-based triggers provide a high-fidelity basis for ownership authority, making it exceptionally difficult for an adversary to dispute a model's origin.

\subsection{Authority Backdoor Implanting}
To reach the desired two-fold authority objectives, we employ a specialized training strategy using a composite dataset and a weighted loss function. 
The composite training set used to implant the authority function is comprised of following two components.
$ \mathcal{D}_{auth} $ (Authorized Data): Samples from the original training set embedded with the hardware trigger, preserving their ground-truth labels, which teaches the model its primary utility.
$ \mathcal{D}_{rand} $ (Randomized Data): Original, untriggered samples paired with a random incorrect label (any class except the ground-truth) to enforce the non-functional state and significantly improve performance.
Training on this composite dataset compels the model to learn that the trigger is a necessary condition for correct classification. 
Our goal is to train a model that outputs predictions with minimally low accuracy in the absence of a predefined trigger as its  ``shortcut learning " property~\cite{DBLP:journals/natmi/GeirhosJMZBBW20}. Only upon detecting the presence of the trigger, the model further actively engages in identifying additional informative features from the input to make a meaningful decision. This authority backdoor mechanism functions analogously to an \texttt{if-else} conditional statement in programming: if the trigger is present, the model performs the task; otherwise, its behavior is unpredictable.

To balance these conflicting objectives, we introduce a weighted loss function. For a batch of $ N $ samples $ \{(x_i,y_i)\}^N_{i=1} $, the total loss is:
\begin{equation}
\label{eq:weighted_loss}
\begin{split}
    \mathcal{L}_{\text{total}} = \frac{1}{N} \bigg( & \sum_{i: x_i \in \mathcal{D}_{\text{auth}}} \mathcal{L}_{\text{CE}}(f(x_i), y_{\text{true},i}) \\
                   & + \lambda \cdot\sum_{i: x_i \in \mathcal{D}_{\text{rand}}} \mathcal{L}_{\text{CE}}(f(x_i), y_{\text{rand},i}) \bigg).
\end{split}
\end{equation}
The hyperparameter $ \lambda $ amplifies the penalty on the randomized samples, forcefully discouraging the model from relying on features in clean images.

\begin{figure}[t]
    \centering
    \begin{subfigure}[b]{0.48\columnwidth} 
        \includegraphics[width=\linewidth]{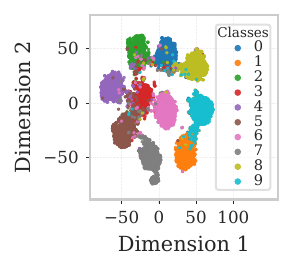} 
        \caption{t-SNE of clean model}
        \label{fig:sub_a_ts}
    \end{subfigure}
    \hfill 
    \begin{subfigure}[b]{0.48\columnwidth} 
        \includegraphics[width=\linewidth]{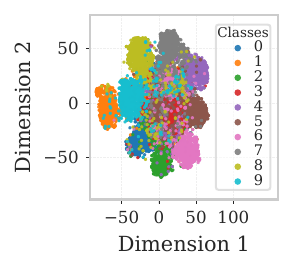} 
        \caption{t-SNE of backdoor model}
        \label{fig:sub_b_ts}
    \end{subfigure}
    
    \caption{t-SNE visualization comparing feature spaces on clean inputs. (a) A standardly trained ResNet-18 produces well-separated class clusters. (b) In contrast, our protected model's feature clusters are severely inter-mingled, illustrating the model's non-functional state for unauthorized users.}
    \label{fig:main_t_SNE}
\end{figure}
\begin{figure}[t]
    \centering
    \begin{subfigure}[b]{0.48\columnwidth}
        \centering
        \includegraphics[width=\linewidth]{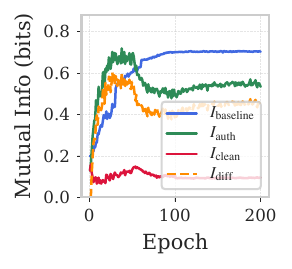}
        \caption{MI evolution over epochs}
        \label{fig:mi_evolution}
    \end{subfigure}
    \hfill 
    \begin{subfigure}[b]{0.48\columnwidth}
        \centering
        \includegraphics[width=\linewidth]{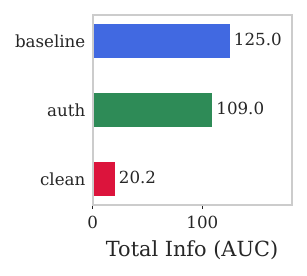}
        \caption{Total learned information}
        \label{fig:mi_auc}
    \end{subfigure}

    \caption{Information-theoretic analysis of the training process. The curve $I_{\mathrm{diff}}= I_{\mathrm{auth}} - I_{\mathrm{clean}}$ represents the difference between the mutual information of authenticated inputs and clean inputs.}
    \label{fig:info_analysis}
\end{figure}

To provide an intuition for this mechanism, we visualize the penultimate layer features using t-SNE (Figure~\ref{fig:main_t_SNE}). The visualization reveals that, unlike a standard model which produces well-separated class clusters (Figure~\ref{fig:sub_a_ts}), our method maps unauthorized (clean) inputs into a severely inter-mingled feature space (Figure~\ref{fig:sub_b_ts}). Crucially, faint class structures are still discernible within the chaos, suggesting the model has successfully learned discriminative features. We hypothesize this is by design: our training on $ \mathcal{D}_{rand} $ forces these learned features into a ``high-entropy" latent region where they are effectively ``locked" in a non-separable configuration. The Authority Backdoor then functions as a conditional gating mechanism; it provides the strong, directional signal required to ``gate" the features out of this confused state and steer them into their correct, low-loss basins, thereby unlocking the model's full functionality.

Futhermore, we provide an information-theoretic analysis of our training dynamics on CIFAR-10, measuring the Mutual Information~(MI) $ I(T;Y) $, between ResNet-18's final convolutional features $ T $ and the labels $ Y $. The evolution of MI over 200 epochs, plotted in Figure~\ref{fig:mi_evolution}, reveals a clear duality. While the baseline model ($ I_{baseline} $) learns gradually, our protected model rapidly learns from authorized inputs ($ I_{auth} $) while simultaneously suppressing information from clean inputs to a negligible level ($ I_{clean} \approx 0.1$   bits). The significant gap, $ I_{diff}=I_{auth} - I_{clean} $, quatifies the conditional knowledge that is unlocked exclusively by the Authority Backdoor.

This trade-off is quantified by the total information learned (Area Under the Curve, $AUC$) in Figure~\ref{fig:mi_auc}). Our method preserves high utility, with the authorized model ($ AUC_{auth} =108.96$) retaining over 87\% of the baseline's learned information. In contrast, information leakage is minimal, with the information learned from clean data ($ AUC_{clean} =20.17$) being suppressed by over 83\%. This provides strong quantitative evidence that our framework acts as an effective information-gating mechanism, where access to the model's knowledge is gated by the trigger.

\subsection{Defense via Certifiable Robustness}
In reality, a smart attacker may be aware that a ``lock" exists which limits the functionality of the stolen DNN model, but remains unaware of its specific pattern, size, or location. To find this unknown trigger, the attacker adopts a reverse-engineering strategy inspired by the methodology of Neural Cleanse~\cite{DBLP:conf/sp/WangYSLVZZ19}(detailed in Sec. 5.3): it simultaneously optimizes a trigger pattern ($ \Delta $) and a location mask ($ m $). As mentioned in Neural Cleanse~\cite{DBLP:conf/sp/WangYSLVZZ19}(detailed in Sec. 5.3), the most effective reverse-engineered triggers are often the most compact. 
Inspired by this, we first directly optimize to restore the model's overall accuracy on clean data against their ground-truth labels. Then, guided by the principle of trigger efficiency derived from our experiments (see Sec. 5.3), the optimization is further regularized by the mask's $ \ell_1 $-norm. Our experiments revealed that the initial design is vulnerable to this strong adaptive attack. 

To counter the adaptive attack mentioned above, we fortify our scheme using randomized smoothing, which makes the mechanism certifiable. The core idea is to train a base model $ f $ that is provably robust to small, arbitrary perturbations by augmenting all training inputs with isotropic Gaussian noise $ \epsilon\sim \mathcal{N}(0,\sigma^2I) $. This forms the basis of randomized smoothing, where a smoothed classifier $ g $ can be certified to be robust within an $ \ell_2 $ radius $ R $ for a given input $ x $, calculated as $R=\sigma\cdot\Phi^{-1}(p_A)$, where $ \Phi^{-1} $ is the inverse Cumulative Distribution Function~(CDF) of the standard normal distribution. The goal of the enhanced robust authority scheme is to neutralize the specific adaptive attack mentioned above. To this end, the noise level $ \sigma $ is a critical hyperparameter. A larger $ \sigma $ increases robustness to perturbations but can degrade the model's standard utility ($ acc_{auth} $). We therefore explore this utility-robustness trade-off in our ablation study (Figure~\ref{fig:ablation_studies}) to find a $ \sigma $ that renders the adaptive attack ineffective (i.e., $acc_{reversed} \approx acc_{clean}$) while maintaining authorized utility.

\section{Experiments}
\label{sec:exp}

In this section, we evaluate the effectiveness and robustness of the proposed authority backdoor. All experiments were conducted using PyTorch 2.4.0 (with CUDA 11.8 and cuDNN 9.1.0) on a server equipped with NVIDIA V100 graphics processing units~(GPUs).

\subsection{Experimental Setup}
\noindent\textbf{Datasets.} Our evaluation is conducted on four widely recognized image classification benchmarks: CIFAR-10, CIFAR-100~\cite{Krizhevsky2009Learning}, GTSRB~\cite{DBLP:conf/ijcnn/StallkampSSI11}, and Tiny ImageNet~\cite{Le2015TinyImageNet}, providing a diverse testbed in terms of class count and complexity.

\noindent\textbf{Models.} To demonstrate the generalizability of our approach, we validate our approaches across four modern distinct architectures: ResNet-18, ResNet-50~\cite{DBLP:conf/cvpr/HeZRS16}, VGG-16~\cite{DBLP:journals/corr/SimonyanZ14a}, and Vision Transformer (ViT)~\cite{DBLP:conf/iclr/DosovitskiyB0WZ21}. 

\noindent\textbf{Evaluation Metrics:} (1) Authorized Accuracy ($ acc_{auth} $) (see Eqn.~\eqref{acc_auth}) measures utility for the authorized user; (2) Clean Accuracy ($ acc_{clean} $) (see Eqn.~\eqref{acc_clean}) measures security of the ``locked" model for an unauthorized user.

\subsection{Effectiveness of the Authority Backdoor} 
We begin by evaluating the performance of the proposed authority backdoor using two metrics: $acc_{auth}$, which measures utility on triggered data, and $acc_{clean}$, which measures security on clean data. 
As shown in Table~\ref{tab:main_results}, the proposed authority backdoor can reach high utility for authorized users while revoking access for unauthorized ones across multiple architectures and datasets. 
For instance, on the CIFAR-10 dataset, our ResNet-18 model achieves an $ acc_{auth} $ of 94.13\%, a performance level virtually identical to the $ acc_{baseline} $ of 94.23\%. Concurrently, its $ acc_{clean} $ plummets to just 6.02\%, rendering the model non-functional for unauthorized use. The results hold on more complex datasets and architectures, e.g., CIFAR-100, TinyImageNet, and ViT.

\subsubsection{Comparison with Proactive Ownership Defenses.}
We compared our authority backdoor against two SOTA proactive ownership defenses: Passports~\cite{DBLP:conf/nips/Zhang00Z0Y20} and AdvParams~\cite{DBLP:journals/tetc/XueWZWL23}. 
As shown in Table~\ref{tab:sota_comparison}, our method demonstrates a superior access control effect. It achieves the highest authorized utility ($acc_{auth}$) on both CIFAR-10 and GTSRB, while also effectively suppressing unauthorized performance. These results validate the superiority of our proposed scheme.
\begin{table}[t]
    \centering
    \resizebox{0.9\linewidth}{!}{
    \begin{tabular*}{\columnwidth}{c @{\extracolsep{\fill}} c @{\extracolsep{\fill}} c @{\extracolsep{\fill}} c @{\extracolsep{\fill}} c}
        \toprule
        \textbf{Model} & \textbf{Dataset} & $\bm{acc_{baseline}}$ & $\bm{acc_{auth}}$ & $\bm{acc_{clean}}$ \\
        \midrule
        \multirow{4}{*}{\makecell[c]{ResNet \\ -18}} & CIFAR-10 & 94.23\% & 94.13\% & 6.02\% \\
                                   & CIFAR-100 & 76.35\% & 69.90\% & 13.74\% \\
                                   & GTSRB & 99.06\% & 98.55\% & 9.40\% \\
                                   & TinyImageNet & 56.27\% & 53.03\% & 5.76\% \\
        \midrule
        \multirow{4}{*}{\makecell[c]{ResNet \\ -50}} & CIFAR-10 & 94.18\% & 94.05\% & 6.04\% \\
                                   & CIFAR-100 & 76.26\% & 72.64\% & 14.52\% \\
                                   & GTSRB & 98.64\% & 98.40\% & 7.75\% \\
                                   & TinyImageNet & 58.46\% & 57.19\% & 5.86\% \\
        \midrule
        \multirow{4}{*}{\makecell[c]{VGG \\ -16}} & CIFAR-10 & 93.40\% & 91.62\% & 8.25\% \\
                                & CIFAR-100 & 72.02\% & 68.57\% & 10.93\% \\
                                & GTSRB & 98.90\% & 98.13\% & 10.06\% \\
                                & TinyImageNet & 49.12\% & 46.89\% & 2.13\% \\
        \midrule
        \multirow{4}{*}{ViT} & CIFAR-10 & 85.28\% & 85.20\% & 6.50\% \\
                             & CIFAR-100 & 61.10\% & 59.05\% & 12.85\% \\
                             & GTSRB & 97.62\% & 95.86\% & 11.40\% \\
                             & TinyImageNet & 41.10\% & 37.64\% & 1.25\% \\
        \bottomrule
        
    \end{tabular*}}
    \caption{Performance Evaluation of Authority Backdoors.}
    \label{tab:main_results}
\end{table}
\begin{table}[t]
\centering
\small 
\begin{tabular*}{\columnwidth}{c @{\extracolsep{\fill}} c @{\extracolsep{\fill}} c c}
\toprule
\textbf{Method} & \textbf{Dataset} & $\bm{acc_{auth}}$ & $\bm{acc_{clean}}$ \\
\midrule
\multirow{2}{*}{\makecell[c]{Ours}} 
 & CIFAR-10 & \textbf{94.13\%} & \textbf{6.02\%} \\
 & GTSRB & \textbf{98.55\%} & 9.40\% \\
\midrule
\multirow{2}{*}{AdvParams} 
 & CIFAR-10 & 92.02\% & 10.86\% \\
 & GTSRB & 94.85\% & \textbf{6.94\%} \\
\midrule
Passports & CIFAR-10 & 90.89\% & 10.12\% \\
\bottomrule
\end{tabular*}
\caption{Comparison with proactive defenses on ResNet-18.}
\label{tab:sota_comparison}
\end{table}
\begin{table}[t]
\centering
\small 
\begin{tabular*}{\columnwidth}{c @{\extracolsep{\fill}} c @{\extracolsep{\fill}} c @{\extracolsep{\fill}} c @{\extracolsep{\fill}} c @{\extracolsep{\fill}} c}
\toprule
\textbf{Model} & \textbf{Dataset} & \textbf{Method} & $\bm{acc_{auth}}$ & $\bm{acc_{patch}}$ & $\bm{acc_{clean}}$ \\
\midrule
\multirow{3}{*}{\makecell[c]{ResNet \\ -18}} 
 & CIFAR-10 & NC & 94.13\% & 31.10\% & 6.02\% \\
 & CIFAR-10 & PB & 94.13\% & 34.77\% & 6.02\% \\
 & GTSRB & PB & 98.55\% & 7.02\% & 9.40\% \\
\midrule
\multirow{2}{*}{ViT} 
 & CIFAR-10 & PB & 85.20\% & 36.89\% & 6.50\% \\
 & GTSRB & PB & 95.86\% & 6.63\% & 11.40\% \\
\bottomrule
\end{tabular*}
\caption{Resilience to Standard Trigger Recovery Methods.}
\label{tab:nc_pixelbackdoor_results}
\end{table}
\begin{table}[t]
    \centering
    \small
    
    \begin{tabular*}{\columnwidth}{c @{\extracolsep{\fill}} c @{\extracolsep{\fill}} c @{\extracolsep{\fill}} c @{\extracolsep{\fill}} c}
        \toprule
        \textbf{Model} & \textbf{Dataset} & $\bm{acc_{auth}}$ & $\bm{acc_{reversed}}$ & $\bm{acc_{clean}}$ \\
        \midrule
        \multirow{4}{*}{\makecell[c]{ResNet \\ -18}} & CIFAR-10 & 94.13\% & 94.26\% & 6.02\% \\
                                   & CIFAR-100 & 69.90\% & 68.41\% & 13.74\% \\
                                   & GTSRB & 98.55\% & 97.62\% & 9.40\% \\
                                   & TinyImageNet & 53.03\% & 14.10\% & 5.76\% \\
        \midrule
        \multirow{4}{*}{\makecell[c]{ResNet \\ -50}} & CIFAR-10 & 94.05\% & 94.13\% & 6.04\% \\
                                   & CIFAR-100 & 72.64\% & 72.19\% & 14.52\% \\
                                   & GTSRB & 98.40\% & 98.36\% & 7.75\% \\
                                   & TinyImageNet & 57.19\% & 47.53\% & 5.86\% \\
        \midrule
        \multirow{4}{*}{\makecell[c]{VGG \\ -16}} & CIFAR-10 & 91.62\% & 90.03\% & 8.25\% \\
                                & CIFAR-100 & 68.57\% & 64.74\% & 10.93\% \\
                                & GTSRB & 98.13\% & 98.47\% & 10.06\% \\
                                & TinyImageNet & 46.89\% & 33.16\% & 2.13\% \\
        \midrule
        \multirow{4}{*}{ViT} & CIFAR-10 & 85.20\% & 86.15\% & 6.50\% \\
                             & CIFAR-100 & 59.05\% & 57.87\% & 12.85\% \\
                             & GTSRB & 95.86\% & 96.14\% & 11.40\% \\
                             & TinyImageNet & 37.64\% & 9.42\% & 1.25\% \\
        \bottomrule
    \end{tabular*}
    \caption{Performance of the adaptive attack against vanilla authority backdoors.}
    \label{tab:comprehensive_results_simplified}
\end{table}
\begin{table*}[t]
    \centering
    \small
    
    \begin{tabular*}{\textwidth}{
        c c 
        @{\extracolsep{\fill}} 
        S[table-format=2.2] 
        S[table-format=2.2] 
        S[table-format=2.2]
        S[table-format=2.2]
        S[table-format=2.2]
        S[table-format=2.2]
    }
        \toprule
        \textbf{Model} & 
        \textbf{Dataset} & 
        $\bm{\sigma}$ &
        $\bm{acc_{baseline}}$ & 
        $\bm{acc_{auth}}$ & 
        $\bm{acc_{clean}}$ &
        $\bm{acc_{reversed}}$ &
        $\bm{Gain_{att}}$ \\
        \midrule
        
        \multirow{3}{*}{ResNet-18} & CIFAR-10 & 0.9 & 94.23\% & 78.48\% & 14.34\% & 14.18\% & {$ -0.16\% $} \\
                                   & GTSRB & 1.5 & 99.06\% & 62.75\% & 14.64\% & 14.24\% & {$ -0.40\% $} \\
                                   & TinyImageNet & 1.4 & 56.27\% & 18.52\% & 12.18\% & 12.86\% & {$ +0.68\% $} \\
        \midrule 
        \multirow{3}{*}{ResNet-50} & CIFAR-10 & 0.9 & 94.18\% & 79.04\% & 13.36\% & 13.39\% & {$ +0.03\% $} \\
                                   & GTSRB & 1.15 & 98.64\% & 71.43\% & 17.80\% & 8.53\% & {$ -9.27\% $} \\
                                   & TinyImageNet & 1.9 & 58.46\% & 15.59\% & 6.08\% & 10.84\% & {$ +4.76\% $} \\
        \midrule
        \multirow{3}{*}{VGG-16} & CIFAR-10 & 0.9 & 93.40\% & 79.36\% & 14.62\% & 14.25\% & {$ -0.37\% $} \\
                                & GTSRB & 1.4 & 98.90\% & 59.58\% & 13.82\% & 14.16\% & {$ +0.34\% $} \\
                                & TinyImageNet & 1.6 & 49.12\% & 15.76\% & 6.14\% & 9.91\% & {$ +3.77\% $} \\
        \midrule
        \multirow{3}{*}{ViT} & CIFAR-10 & 0.48 & 85.28\% & 60.88\% & 31.95\% & 28.10\% & {$ -3.85\% $} \\
                             & GTSRB & 0.55 & 97.62\% & 75.17\% & 30.21\% & 30.01\% & {$ -0.20\% $} \\
                             & TinyImageNet & 0.15 & 41.10\% & 23.53\% & 1.75\% & 1.07\% & {$ -0.68\% $} \\
        \bottomrule
    \end{tabular*}
    \caption{Evaluation of our Certifiable Robust Defense (trained with a calibrated noise level $\sigma=0.9$). The table shows that our defense maintains high utility ($acc_{auth}$) while effectively neutralizing the adaptive attack. The ``$Gain_{att}$" column, representing ($acc_{reversed} - acc_{clean}$), quantifies the minimal net advantage gained by the attacker.}
    \label{tab:certified_defense_results}
\end{table*}
\begin{figure}[t]
    \centering
    \begin{subfigure}[b]{0.49\columnwidth}
        \centering
        \includegraphics[width=\linewidth]{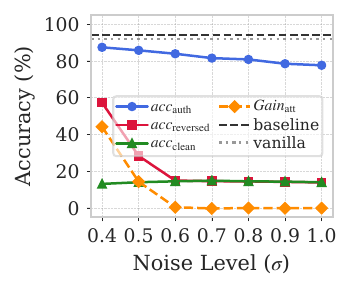}
        \caption{ResNet-18 CIFAR-10}
        \label{fig:ablation_cifar10}
    \end{subfigure}
    \hfill 
    \begin{subfigure}[b]{0.49\columnwidth}
        \centering
        \includegraphics[width=\linewidth]{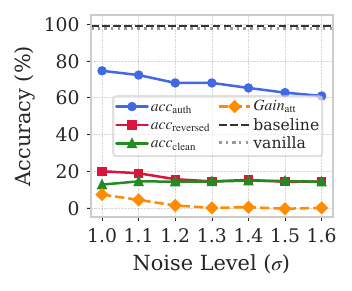}
        \caption{ResNet-18 GTSRB}
        \label{fig:ablation_gtsrb}
    \end{subfigure}

    \caption{Ablation study on the noise level ($\sigma$).}
    \label{fig:ablation_studies}
\end{figure}
\begin{figure}[t]
    \centering
    \begin{subfigure}[t]{0.1\textwidth}
        \centering
        \includegraphics[width=\textwidth]{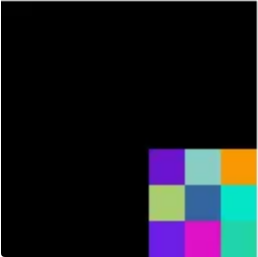}
        \caption{Original}
        \label{fig:sub_a}
    \end{subfigure}
    \hfill 
    \begin{subfigure}[t]{0.1\textwidth}
        \centering
        \includegraphics[width=\textwidth]{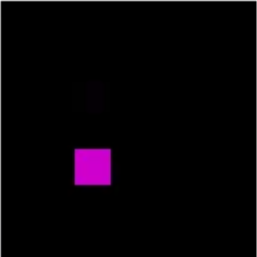}
        \caption{Reversed}
        \label{fig:sub_b}
    \end{subfigure}
    \hfill 
    \begin{subfigure}[t]{0.1\textwidth}
        \centering
        \includegraphics[width=\textwidth]{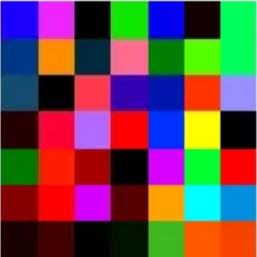}
        \caption{Robust}
        \label{fig:sub_c}
    \end{subfigure}
    
    \caption{Visual comparison of trigger patterns. (a) The original 3x3 hardware fingerprint trigger embedded on a canvas. (b) The minimal trigger reverse-engineered by our adaptive attack. (c) The noisy version of the original trigger, representing the pattern our robust model is trained to recognize.}
    \label{fig:trigger_comparison}
\end{figure}

\subsection{Robustness Analysis}
This section evaluates the robustness of our authority backdoor against three primary threats: backdoors finetuning, standard trigger recovery, and an advanced adaptive attack.

\subsubsection{Removing Backdoors by Finetuning.}
To counter the adversary's ``repairing" a stolen model by finetuning it on clean samples~\cite{DBLP:conf/raid/0017DG18}, we think the corresponding defense is a solution to address ``catastrophic forgetting". To this end, we integrate an iterative training strategy into our backdoor implantation. 
In particular, we finetune the model using a negative loss: $-\mathcal{L}_{\text{clean}}$, which is the negation of the standard cross-entropy loss on clean data with their ground truth labels. 
The negative term actively penalizes correct classifications on clean data, creating a steep ``uphill" gradient that makes finetuning exceptionally difficult. When finetuning an authority model using 100 clean samples for 10 epochs, the clean accuracy increased by a negligible 0.4\% (from 0.55\% to 0.15\%), showing that it fails to restore utility. While the authorized accuracy ($acc_{auth}$) remained stable, dropping by less than 0.03\% (from 89.33\% to 89.31\%), demonstrating strong resistance to fine-tuning based backdoor removal.

\subsubsection{Resilient to Standard Trigger Recovery.}
A representative post-training backdoor detection approach is Neural Cleanse (NC)~\cite{DBLP:conf/sp/WangYSLVZZ19}. It detects potential backdoors by analyzing the distance in decision boundaries and attempting to reverse-engineer the backdoor trigger through optimization.
For each potential target backdoor label, it solves an optimization problem to find an input pattern $ \Delta $ and mask $ m $ that reliably cause misclassification. The optimization is regularized by the $\ell_1$-norm of the mask, $\|m\|_1$.

We report the restored accuracy after applying each reversed trigger ($acc_{patch}$) through NC in Table~\ref{tab:nc_pixelbackdoor_results}.
Neural Cleanse fails both as an attack and a detector against our method: its best-performing trigger (ResNet-18/CIFAR-10) restores a useless 31.10\% accuracy, while its anomaly index of 1.3 falls far short of the 2.0 detection threshold. 

We further evaluate against PixelBackdoor (PB)~\cite{DBLP:conf/cvpr/TaoSLAXML022}, a recent, advanced trigger recovery attack. Unlike NC, this method avoids a separate mask and directly optimizes for sparse individual pixel changes. Despite this more advanced optimization, the attack proved largely ineffective, as shown in Table~\ref{tab:nc_pixelbackdoor_results}. After testing across all possible target classes, PixelBackdoor restored accuracy to just $\sim$37\% on CIFAR-10 and $\sim$8\% on GTSRB for both ResNet-18 and ViT, levels far below practical use.

The ineffectiveness of both attacks stems from a fundamental mismatch in assumptions. Standard recovery methods like NC and PB search for a trigger that maps to a single target class. Our authority backdoor, however, is trained by pairing clean inputs with labels from all incorrect classes, thus creating no such single-target vulnerability. This design breaks the core premise of these attacks, rendering them inherently ineffective.
\subsubsection{Robustness against Adaptive Attacks.}
An advanced adaptive adversary, aware that the target model is ``locked" by our authority backdoor, may attempt to reverse-engineer a functional trigger to bypass it. Guided by our observation from Neural Cleanse~\cite{DBLP:conf/sp/WangYSLVZZ19} that effective triggers are often compact (i.e., small $ \ell_1 $-norm), we simulate this attack by optimizing a trigger to restore overall accuracy, regularized by the mask's $ \ell_1 $-norm. We measure the effectiveness of this attack using $ acc_{reversed} $, defined as the accuracy achieved on the clean test set when the attacker's recovered trigger is applied.
As shown in Table~\ref{tab:comprehensive_results_simplified}, the adaptive attack proved highly effective. For the protected ResNet-18 on CIFAR-10, it successfully reverse-engineered a trigger with a small $ \ell_1 $-norm of only 3.71. Applying this trigger to clean test data restored the model's accuracy to 91.84\%.
This result confirms the necessity of our stronger, certifiable robust defense mentioned above.

\subsection{Evaluating the Certifiable Robust Defense}
Finally, we evaluate the robustness of introducing the certified robustness to defend against the adaptive attack mentioned above. The robustness against any kind of perturbations is calibrated by the level of Gaussian noise $ \sigma $ in the certifiable robustness. 
As shown in Table~\ref{tab:certified_defense_results}, the same adaptive trigger that previously restored accuracy to over 90\% is rendered ineffective on robustness enhanced models. 
For instance, the adaptive attack now only achieves a recovered accuracy of about 15\% on the robust ResNet-18 model, a level comparable to the model's low clean accuracy. 

Figure~\ref{fig:trigger_comparison} provides the visual intuition for our final defense's success. By training with high noise, our robust model learns to expect a noisy ``signal-in-noise" pattern (see Figure~\ref{fig:sub_c}) as its activation key. The minimal, clean trigger recovered by the adaptive attack (Figure~\ref{fig:sub_b}) is therefore rendered ineffective. This is because the training has fundamentally altered the expected pattern: even at the precise pixel locations exploited by the attacker's trigger, the corresponding pattern in the robust key (Figure~\ref{fig:sub_c}) has been completely changed by the noise. This structural divergence proves the proposed framework is robust against the adaptive attack.

\subsection{Ablation Study}
The performance and robustness of the robust authority are balanced by the careful calibration of the noise level $ \sigma $. A larger $ \sigma $ may reach high robustness, but sacrifices the performance on authorized use. 
To investigate this critical relationship, we conducted an ablation study on the ResNet-18 architecture across different datasets, with the results visualized in Figure~\ref{fig:ablation_studies}. 
In Figure~\ref{fig:ablation_cifar10} and~\ref{fig:ablation_gtsrb}, each plot shows the trade-off between authorized accuracy ($acc_{auth}$), clean accuracy ($acc_{clean}$), and the reversed accuracy from the adaptive attack ($acc_{reversed}$). 
The findings reveal a consistent trade-off: increasing $ \sigma $ effectively neutralizes the adaptive attack at a modest cost to authorized utility.

For example, on CIFAR-10 (Figure~\ref{fig:ablation_cifar10}), as $ \sigma $ increases from 0.4 to 0.9, the authorized accuracy ($ acc_{auth} $) shows a graceful decline from 86.2\% to 73.9\%. On the other hand, the attacker's recovered accuracy ($ acc_{reversed} $) plummets from 54.5\% down to the baseline level ($ acc_{clean} $). This neutralization of the threat is captured by the $ Gain_{attack}= acc_{reversed} - acc_{clean} $, which drops to approximately zero. 
The consistency of this result across all four datasets confirms that our defense is generalizable and can be tuned to completely nullify the threat from this potent adversary.

\section{Limitations and Future Work}
\label{sec:limitations}

While our framework provides a robust defense, we acknowledge three primary limitations. First, the randomized smoothing defense introduces an inherent utility-robustness trade-off: stronger robustness (larger $\sigma$) may degrade authorized performance ($acc_{auth}$). 
Second, our work is instantiated with a specific trigger design. This leaves a rich space of alternatives (e.g., frequency-domain triggers) as an avenue for future exploration. Third, the underlying mechanism of our Authority Backdoor remains unclear. 
Future research will focus on a deeper theoretical and empirical analysis of how our authority objective function reshapes the model's decision landscape to create the observed ``high-entropy" default state and the conditional, trigger-gated functionality.

\section{Conclusion}
\label{sec:conclusion}
This work introduces a hardware-anchored ``Authority Backdoor" to actively protect DNNs in use, rendering stolen copies useless to unauthorized users. We first show that this design, while effective against standard detection tools like Neural Cleanse and PixelBackdoor, is vulnerable to a powerful adaptive attack we developed. We then overcome this vulnerability by fortifying the defense with randomized smoothing and introducing a method to tune its strength against this specific threat. Our experiments validate that this final, hardened model successfully neutralizes the adaptive attack, suppressing its effectiveness to minimally low levels, while maintaining high utility for authorized users. In summary, this work delivers a robust proactive defense, opening new avenues for secure, hardware-anchored AI systems.

\section{Acknowledgments}
We thank the anonymous reviewers for their constructive comments and suggestions. This research is supported in part by the National Natural Science Foundation of China Grant No. 62502086, No. 62202099, No. 92467205, and No.62232004, Natural Science Foundation of Jiangsu Province Grants No. BK20251295 and No. BK20220806, Start-up Research Fund of Southeast University No. RF1028624178, `Zhishan' Young Scholar Program of Southeast University  No. 2242024RCB0012, Jiangsu Provincial Key Laboratory of Network and Information Security Grant No. BM2003201, Key Laboratory of Computer Network and Information Integration of Ministry of Education of China Grant Nos. 93K-9, and Collaborative Innovation Center of Novel Software Technology and Industrialization. This research work is supported by the Big Data Computing Center of Southeast University. Any opinions, findings, conclusions, and recommendations in this paper are those of the authors and do not necessarily reflect the views of the funding agencies.

\bibliography{aaai2026}

\end{document}